\documentclass[journal]{IEEEtran}
\usepackage{makecell}
\usepackage{multirow}
\usepackage{svg}
\usepackage{epstopdf}
\usepackage{float}
\usepackage{cite}
\usepackage{bm}
\usepackage{pifont}
\usepackage{bbding}
\usepackage{amsmath}
\usepackage{amssymb}
\usepackage{mathrsfs,balance}
\usepackage{algorithm}
\usepackage{algorithmic}
\usepackage{color}
\usepackage{amsthm}
\usepackage{caption}
\usepackage{graphicx}
\usepackage{subcaption}
\usepackage{booktabs}
\usepackage{romannum}
\usepackage{xcolor}
\usepackage{hyperref} 

\begin{document}
\pagenumbering{arabic} 
\title{Intelligent Disruption: Undetectable Attacks on Wireless Autoencoders}
\author{Han~Jiang,
		Jifa~Zhang,
		Hu~Jin,~\IEEEmembership{Senior Member,~IEEE},
        Nan~Zhao,~\IEEEmembership{Senior Member,~IEEE},
        Mingqian~Liu,~\IEEEmembership{Member,~IEEE},
       and Yunfei~Chen,~\IEEEmembership{Fellow,~IEEE}
\thanks{H.~Jiang, J.~Zhang and N.~Zhao are with the School of Information and Communication Engineering, Dalian University of Technology, Dalian, Liaoning, P. R. China (email: hanjiang@mail.dlut.edu.cn, jifazhang@mail.dlut.edu.cn, zhaonan@dlut.edu.cn).}
\thanks{ H.~Jin is with the School of Electrical Engineering, Hanyang University, Ansan 15588, South Korea (e-mail: hjin@hanyang.ac.kr).}
\thanks{M.~Liu is with the State Key Laboratory of Integrated Services Networks, Xidian University, Xi’an 710071, China (e-mail: mqliu@mail.xidian.edu.cn).}
\thanks{Y.~Chen is with Department of Engineering, University of Durham, Durham, UK DH1 3LE (e-mail: Yunfei.Chen@durham.ac.uk).}
\thanks{N.~Zhao is the corresponding author.}
}
\maketitle
\begin{abstract}
Adversarial attacks can degrade the legitimate decision performance in wireless autoencoder communications. However, in complex scenarios with multiple adversaries, the cumulative leakage interference (CLI) caused by the multiple parallel attacks increases the chance of detecting the attacks, while dynamical environments also make the fixed attack strategies difficult to have stable effectiveness. To jointly enhance the undetectability, aggressivity and adaptability of adversarial attacks, we propose a deep learning based intelligent attack framework. Specifically, considering the CLI caused by the multiple parallel attacks, a deep neural network based transmit power control is established to reduce the interference leakage by regulating the transmit power of these adversaries, thereby improving the undetectability. Furthermore, to enhance the attack effectiveness and stability in the dynamic environment, the conditional generative adversarial attack is further developed. The generator takes the attack channel information as the conditional input to produce the perturbating signals to mislead the discriminator by making the attacked received signals resemble the clean received signals, while the discriminator distinguishes between the two under the same condition. Through the adversarial training, the generator can learn to create adaptive perturbating signals with enhanced attack performance.  
Simulation results demonstrate that the proposed framework outperforms benchmarks in terms of attack undetectability, aggressivity and adaptability.

\end{abstract}
\begin{IEEEkeywords}
Adversarial attacks, conditional generative adversarial network,  deep neural network, dynamic channel, wireless autoencoder.
\end{IEEEkeywords}
\IEEEpeerreviewmaketitle

\section{Introduction}
Conventional communications generally adopt a modular design, consisting of multiple functional blocks such as modulation and demodulation, channel encoding and decoding, and channel estimation and detection, each of which is usually designed and optimized independently to improve the performance \cite{spereate}. Although this modular design has been widely adopted in practice, the lack of joint optimization across modules seriously limits the communication performance in the complex wireless environment characterized by multipath propagation, time-variance and multi-user interference \cite{modular_limitation}.

Deep learning (DL), with its powerful nonlinear modeling and hierarchical feature representation capabilities, provides a promising paradigm for the end-to-end design and optimization of wireless communications \cite{DL_autoencoder,cui2025overview}. In this context, the DL-based autoencoder communications have attracted increasing attention \cite{review}. Different from the conventional modular design, the transceiver chain in the DL-based autoencoder communications is implemented as an end-to-end autoencoder architecture \cite{introduction,Data_Rate}.
Specifically, the encoder acts as the transmitter, mapping the original message into the transmitted signal, while the decoder serves as the receiver, recovering the original message from the received signal affected by the noise, fading and interference \cite{Fitting}. By jointly training the encoder and decoder, the DL-based autoencoders can adaptively learn the effective signal representation and transceiver strategy, thereby enhancing the communication performance in the complex wireless environment \cite{adv_autoencoder}.

Despite these advantages, the DL-based autoencoder communications are vulnerable to the adversarial attacks due to the broadcast nature of wireless. \cite{Toward}. 
In such attacks, carefully designed perturbations can be added to the original inputs to mislead the autoencoders for erroneous outputs \cite{ren2020adversarial,perturbation}. 
For example, Sadeghi and Larsson in \cite{Physical} proposed an input-agnostic adversarial attack  against an end-to-end DL-based autoencoder system, revealing that the adversarial attacks can be more destructive than jamming attacks.
In \cite{Performance}, Albaseer \textit{et al.} investigated the vulnerability of DL-based autoencoder systems over Rayleigh channels in both the fast-fading and slow-fading scenarios, showing that the fast-fading channels are more susceptible to the adversarial attacks than the slow-fading ones.
Considering a more realistic channel model, Jiang \textit{et al.} in \cite{VSAA} proposed a variable step-size double-iteration attack against the end-to-end DL-based autoencoder, where the channels from both the transmitter and adversary towards the receiver were taken into account. 
In \cite{RIS_Assisted}, Son \textit{et al.} investigated the vulnerability of double reconfigurable intelligent surfaces assisted autoencoder systems in the finite-scattering environment, and proposed the projected gradient descent based adversarial attacks to effectively degrade the reliability.

\begin{table*}[htbp]
	\centering
	\caption{Comparison between our work and the related works.}
	\label{Motivation_comparison}
	\renewcommand{\arraystretch}{1.2} 
	\begin{tabular}{c|c|c|c|c|c|c}
		\hline
		\hline
		\textbf{Reference} & \textbf{Legal Receiver} & \textbf{Antenna Setting} & \textbf{Adversary} & \textbf{CLI} & \textbf{Attack Channel} & \textbf{Adaptability to Dynamic Attack Channels} \\
		\hline
		\cite{Physical}                  & Single    &  SISO  & Single     & \ding{55} & \ding{55} & \ding{55} \\
		\cite{Performance}               & Single    &  SISO  & Single     & \ding{55} & \ding{55} & \ding{55} \\
		\cite{VSAA}              & Single    &  SISO  & Single     & \ding{55} & \ding{51} & \ding{55} \\
		\cite{RIS_Assisted}              & Single    &  MIMO  & Single     & \ding{55} & \ding{55} & \ding{55} \\
		\cite{Channel_RobustSISO6}       & Single    &  SISO  & Single     & \ding{55} & \ding{55} & \ding{51} \\
		\cite{Reinforcement}             & Multiple    &  SISO  & Single     & \ding{55} & \ding{55} & \ding{55} \\
		\cite{Channel_Aware}             & Multiple    &  SISO  & Single     & \ding{55} & \ding{51} & \ding{55} \\
		\cite{Defending}                 & Multiple    &  MISO  & Single     & \ding{55} & \ding{55} & \ding{55} \\
		\cite{Downlink}                  & Multiple        & MISO & Single   & \ding{55} & \ding{55} & \ding{55} \\
		\cite{Frequency_Selective}       & Single    &  SISO  & Single     & \ding{55} & \ding{55} & \ding{55} \\
		\cite{Robust}                    & Single    &  SISO  & Single     & \ding{55} & \ding{55} & \ding{55} \\
		\cite{SISO3}                     & Single    &  SISO  & Single     & \ding{55} & \ding{55} & \ding{55} \\
		\cite{SISO7}                     & Single    &  SISO  & Single     & \ding{55} & \ding{55} & \ding{55} \\
		\cite{Multi_Objective}           & Single    &  SISO  & Single     & \ding{55} & \ding{55} & \ding{55} \\
		\cite{Multiple_Antennas}         & Single    &  SISO  & Single     & \ding{55} & \ding{51} & \ding{55} \\
		\cite{Power_Control}             & Multiple    &  SISO  & Single     & \ding{55} & \ding{51} & \ding{55} \\
		\cite{Universal_Adversarial_Attacks} & Multiple    & MISO & Single   & \ding{55} & \ding{55} & \ding{55} \\
		
		\textbf{This work}               & Multiple   & MISO & Multiple & \ding{51} & \ding{51} & \ding{51} \\
		\hline
		\hline 
	\end{tabular}
	\vspace{1mm}
\end{table*}

Nevertheless, the existing attack methods still suffer from some practical limitations. First, the considered scenarios remain simplified. Although various settings have been considered, including the single input single output (SISO) with a single transmitter and a single receiver \cite{Channel_RobustSISO6,MitigationSISO4},  the SISO with a single transmitter and multiple receivers \cite{Reinforcement,Channel_Aware}, the multiple input multiple output (MIMO) with a single transmitter and a single receiver \cite{RIS_Assisted} and the MIMO with a single transmitter and multiple receivers \cite{Defending,Downlink}, the existing studies mainly focus on the single-adversary case. Such setting may be insufficient to reflect the real adversarial environment, where multiple multi-antenna adversaries may launch the targeted attacks via beamforming to the legitimate receivers simultaneously. Meanwhile, due to the energy leakage and the openness of wireless environment, the interference leakage from the multiple parallel attack links may accumulate at the receivers and form the non-negligible cumulative leakage interference (CLI). Such CLI is more likely to be detected by receivers, thereby exposing the presence of adversaries and causing the failure of attacks.

On the other hand, most of the above attack methods exhibit limited adaptability to the time-varying channels. In practical adversarial scenarios, the perturbations designed at the adversary side have to go through the attack channel from the adversary to receiver before attacking the receiver. During this process, the channel fading may greatly degrade the effectiveness of adversarial attacks. However, the existing studies primarily focus on modeling the communication channel between the transmitter and receiver, while neglecting the attack channel \cite{Frequency_Selective, Energy_Efficient, Siameseatttchannel}. Moreover, the factors such as target mobility may lead to rapid changes in attack channels, making it challenging to maintain stable attack performance in these dynamic environments.


Motivated by these, we investigate the adversarial attacks with multi-adversary. Compared with the single-adversary setting, the multi-adversary one better reflects the practical adversarial environment. In this multi-adversary environment, the parallel attacks inevitably introduce CLI, which increases the risk of exposure. Moreover, since the attack channels vary rapidly due to the target mobility, the adversarial attacks should adapt to the channel changing to maintain stable performance. Table \ref{Motivation_comparison} compares our work with the related publications, and the main contributions are listed as below.

\begin{itemize}
	\item We propose an intelligent attack framework against the MISO DL-based autoencoder with multiple receivers and adversaries, which jointly improves the undetectability, aggressivity and adaptability of adversarial attacks in the practical adversarial environment. 
	
	\item Considering the CLI caused by the multiple parallel attack links, a deep neural network (DNN) based power control is established to mitigate the CLI at each receiver by regulating the transmit power of multiple adversaries, thereby enhancing the undetectability of attacks.
	
	\item To further improve the attack effectiveness in the dynamic adversarial environment, the conditional generative adversarial network (cGAN) based adversarial attack is further developed. By exploiting the attack channel information, the generator can produce the adaptive perturbating signals with enhanced aggressivity.
\end{itemize}

The rest of this paper is organized as follows. Section \Romannum{2} introduces the system model. DNN based power control and cGAN based adversarial attack are presented in Section \Romannum{3} and Section \Romannum{4}, respectively. Section \Romannum{5} provides the simulation results, and the conclusion is given in Section \Romannum{6}.

\textit{Notation:} $\sqrt{\cdot}$, $(\cdot)^{T}$ and $(\cdot)^{H}$ stand for square root, transpose and conjugate transpose, respectively. $\mathcal{CN}\{\mu, \boldsymbol{\Phi}\}$ denotes the circularly symmetric complex Gaussian distribution with mean $\mu$ and covariance matrix $\boldsymbol{\Phi}$. ${\rm arg} (\cdot)$ denotes the angle of a complex number. $\|\cdot\|_2$ represents the $\ell_2$ norm. $\mathbb{E}[\cdot]$ stands for the mathematical expectation.

\begin{figure}[t]
	\centering
	\includegraphics[width=3.4in]{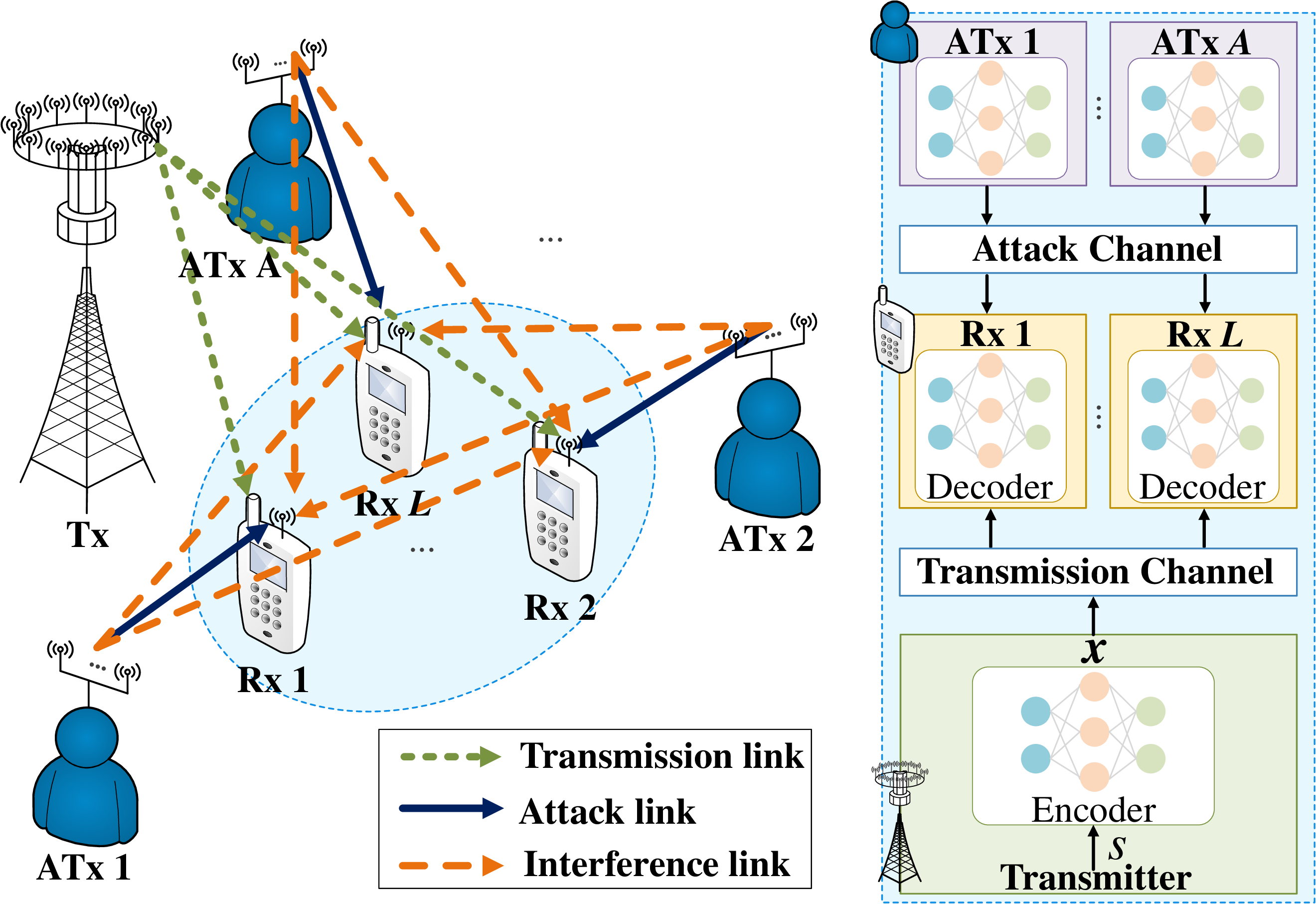} 
	\caption{Multi-adversary MISO DL-based autoencoder system.}
	\label{system_model}
\end{figure}
\section{System Model}
Consider a multi-adversary and multi-receiver MISO DL-based autoencoder system consisting of a transmitter with $M$ antennas, $L$ single-antenna receivers and $A$ $N$-antenna adversaries, where $A=L$ is assumed for the target attack, as shown in Fig. \ref{system_model}. The transmitter is composed of an encoder, and each receiver is equipped with a decoder, where both the encoder and decoder are implemented via neural networks. Each adversary attacks its targeted receiver (TR) via beamforming. However, due to the energy leakage and the openness of wireless environment, each adversary inevitably introduces interference to the non-targeted receivers. 
We assume that the transmitter, receivers and adversaries are synchronized during the transmission, and we adopt an information-rich adversary model for the worst-case vulnerability analysis.
\subsection{Transmission Model}
In this paper, the wireless channels are modeled as Rayleigh fading over the transmission period, which is evenly divided into $Q$ time slots. 
Denote the original input message of the transmitter as $s$, where $s \in \mathcal{U}=\{1,2,\cdots,U\}$, and $U=2^k$ represents the dimension of $\mathcal{U}$ with $k$ being the number of bits per message. 
The message is then passed to the encoder, which applies a transformation $\mathcal{F}:\mathcal{U} \rightarrow \mathbb{C}^{B}$ to form the transmitted signal $\boldsymbol{x} = \mathcal{F}(s) =  [x_1,...,x_b,...x_B]^T \in \mathbb{C}^{1 \times B}$, where $B$ denotes the length of transmitted signal and $Q$ is equal to $B$ to guarantee the complete transmission.
In each time slot, a single symbol $x_b$ is broadcast to all receivers via the channels, where $b \in \{1,...,B\}$. Assume that the channels remain constant within each time slot but vary independently across slots.

Without the attack,  the received signal at the $l$-th receiver in the $b$-th time slot can be expressed as 
\begin{equation}\label{channel_output}
y_{l,b} = \sqrt{P_x}\boldsymbol{h}_{l,b}^{ H} \boldsymbol{w}_x x_{b} +n_{l,b},
\end{equation}
where $l \in\{1,...,L\}$, $P_x$ denotes the  transmit power of transmitter, $\boldsymbol{h}_{l,b} = [h_{l,b}^1,...,h_{l,b}^M] \in \mathbb{C}^{M \times 1}$ and $\boldsymbol{w}_x \in \mathbb{C}^{M \times 1} $ denote the channel and beamforming vectors, respectively, where $\boldsymbol{h}_{l,b} \sim \mathcal{CN}(\boldsymbol{0}_M, \boldsymbol{I}_M)$, $\|\boldsymbol{w}_{x}\|^2  = 1$, and $n_{l,b} \sim \mathcal{CN}\left( 0,\sigma^2 \right)$ is the additive white Gaussian noise (AWGN) in the $b$-th time slot with $\sigma^2 = N_0 / (RE)$, where $E$ denotes the energy per bit and $R$ is the transmission rate in bits per channel use. 
Finally, the decoder at the $l$-th receiver decodes $\boldsymbol{y}_l$ using the transformation $\mathcal{G}:\mathbb{C}^{B} \rightarrow \mathcal{U}$ to estimate the original message as
\begin{equation}\label{shat}
\hat{s}_l = \mathcal{G}(\boldsymbol{y}_l),
\end{equation}
where $\boldsymbol{y}_l = [y_{l,1},...,y_{l,b},...,y_{l,B}]^T \in \mathbb{C}^{1 \times B}$ denotes the received signal at the $l$-th receiver during $Q$ time slots.

\subsection{Attack Model}
With the attack,  each adversary transmits a carefully designed perturbating signal over the attack channel to confuse the decoder at its TR into making errors in the estimation. 
Let $\boldsymbol{h}_{l,b}^{a} \in \mathbb{C}^{N \times 1}$ represent the attack channel vector between the $a$-th adversary and the $l$-th TR in the $b$-th time slot, $a \in\{1,...,A\}$, given by
\begin{equation}\label{ha}
\boldsymbol{h}_{l,b}^{a} \sim \mathcal{CN}(\mathbf{0}_N, \boldsymbol{R}_{l,b}^a),
\end{equation}
with
\begin{equation}\label{haa}
\boldsymbol{R}_{l,b}^a = \beta_{l,b}^a \tilde{\boldsymbol{R}}_{l,b}^a,
\end{equation}
where $\beta_{l,b}^a$ characterizes the large-scale fading of the attack channel between the $a$-th adversary and the $l$-th TR in the $b$-th slot as  
\begin{equation}\label{beta}
\beta_{l,b}^a = 10^{\Gamma - 10\delta{\rm log}_{10} \left(  \frac{|o_{l,b} - o_{a,b}|}{d_0}  \right)},
\end{equation}
where $\Gamma$ represents the gain at a reference distance of $d_0$, $\delta$ denotes the pathloss exponent, $o_{l,b} = c_{l,b} + i s_{l,b}\in \mathbb{C}$ and $o_{a,b} = c_{a,b} + i s_{a,b}\in \mathbb{C}$ denote the positions of the $l$-th receiver and the $a$-th adversary in the $b$-th time slot, respectively. In \eqref{haa}, $\tilde{\boldsymbol{R}}_{l,b}^a \in \mathbb{C}^{N \times N}$ represents the normalized spatial correlation matrix between the $a$-th adversary and the $l$-th TR in the $b$-th time slot, which can be calculated via the local scattering model, considering the phase difference and antenna spacing. To improve the computational efficiency, the small-angle approximation is utilized. Denote the spatial correlation between the $u$-th and $v$-th antennas in the $b$-th time slot as 

{\small
\vspace{-3.2mm}
\begin{equation}\label{R}
\tilde{\boldsymbol{R}}_{(l,b),(u,v)}^a = {e}^{j2\pi d_{u,v} {\rm{sin}}(\theta_b)} \cdot {e}^{-\frac{{(ASD \cdot \pi/180)}^2}{2} \cdot (2\pi d_{u,v} {\rm{cos}}(\theta_b) )^2},
\end{equation}
}where $ASD$ denotes the angular standard deviation in angles, and $d_{u,v}$ represents the antenna spacing between the $u$-th and $v$-th antennas of the $a$-th adversary in wavelengths, $u,v\in \{1,2,...,N\}$. $\theta_b$ denotes the angle of arrival in radians in the $b$-th time slot as
\begin{equation}\label{theta}
 \theta_b = {\rm{arg}}( o_{l,b} - o_{a,b}) ={\rm{arctan}} \left(\frac{s_{l,b} - s_{a,b}}{c_{l,b}-c_{a,b}}\right).
\end{equation}
Furthermore, to ensure the consistency between the large-scale fading scalar and the spatial correlation matrix with respect to the propagation characteristics, we enforce
\begin{equation}
\beta_{l,b}^a = \frac{1}{N} {\rm tr}(\boldsymbol{R}_{l,b}^a),
\end{equation}
so that the trace of $\boldsymbol{R}_{l,b}^a$ reflects the average gain of attack channel between the $a$-th adversary and the $l$-th TR.
Denote the perturbating signal launched by the $a$-th adversary towards the $l$-th TR during the transmission as $\boldsymbol{\vartheta}_l^a = [\vartheta_{l,1}^a,...,\vartheta_{l,b}^a,...,\vartheta_{l,B}^a]^T \in \mathbb{C}^{1 \times B}$, where $\vartheta_{l,b}^a$ denotes the perturbating signal in the $b$-th time slot. Then, the perturbed received signal at the $l$-th TR in the $b$-th time slot \eqref{channel_output} can be changed into 
\begin{equation}
\begin{aligned}
y_{l,b}^a &= \underbrace{\sqrt{P_x}\boldsymbol{h}_{l,b}^{H} \boldsymbol{w}_x x_{b}}_{\text{Desired signal}} 
+ \underbrace{\sqrt{P_{a,b}}(\boldsymbol{h}_{l,b}^a)^{ H} \boldsymbol{w}_{a,b} \vartheta_{l,b}^a}_{\text{Adversarial attack}} \\
&\quad + \underbrace{\sum_{j=1, j \neq a}^{A} \sqrt{P_{j,b}}(\boldsymbol{h}_{l,b}^j)^{ H} \boldsymbol{w}_{j,b} \vartheta_{l,b}^j}_{\text{CLI}} 
+ \underbrace{n_{l,b}}_{\text{AWGN}},
\end{aligned}
\label{channel_output_attack}
\end{equation}
where $P_{a,b}$ denotes the transmit power of the $a$-th adversary in the $b$-th time slot, and $\boldsymbol{w}_{a,b} = [w_{a,b}^1,...,w_{a,b}^n,...,w_{a,b}^N] \in \mathbb{C}^{N\times 1}$ denotes the  beamforming vector determining the spatial directivity of transmission with $\|\boldsymbol{w}_{a,b}\|^2  = 1$. The calculation of $\boldsymbol{w}_{a,b}$ relies on the estimated vector $\boldsymbol{\hat h}_{l,b}^{a}$ via the minimum mean square error (MMSE) combining, as adopted in \cite{Channel_Aware}. 
After the transmission, the received signal and the estimation of the $l$-th TR can be obtained as $\boldsymbol{y}_l^a = [y_{l,1}^a,...,y_{l,b}^a,...,y_{l,B}^a]^T \in \mathbb{C}^{1 \times B}$ and $\hat{s}_l = \mathcal{G}(\boldsymbol{y}_l^a)$, respectively.

\section{DNN Based Power Control to Mitigate CLI}
\begin{figure*}[htbp]
	\centering
	\includegraphics[width=5.5in, trim=15 135 12 120, clip]{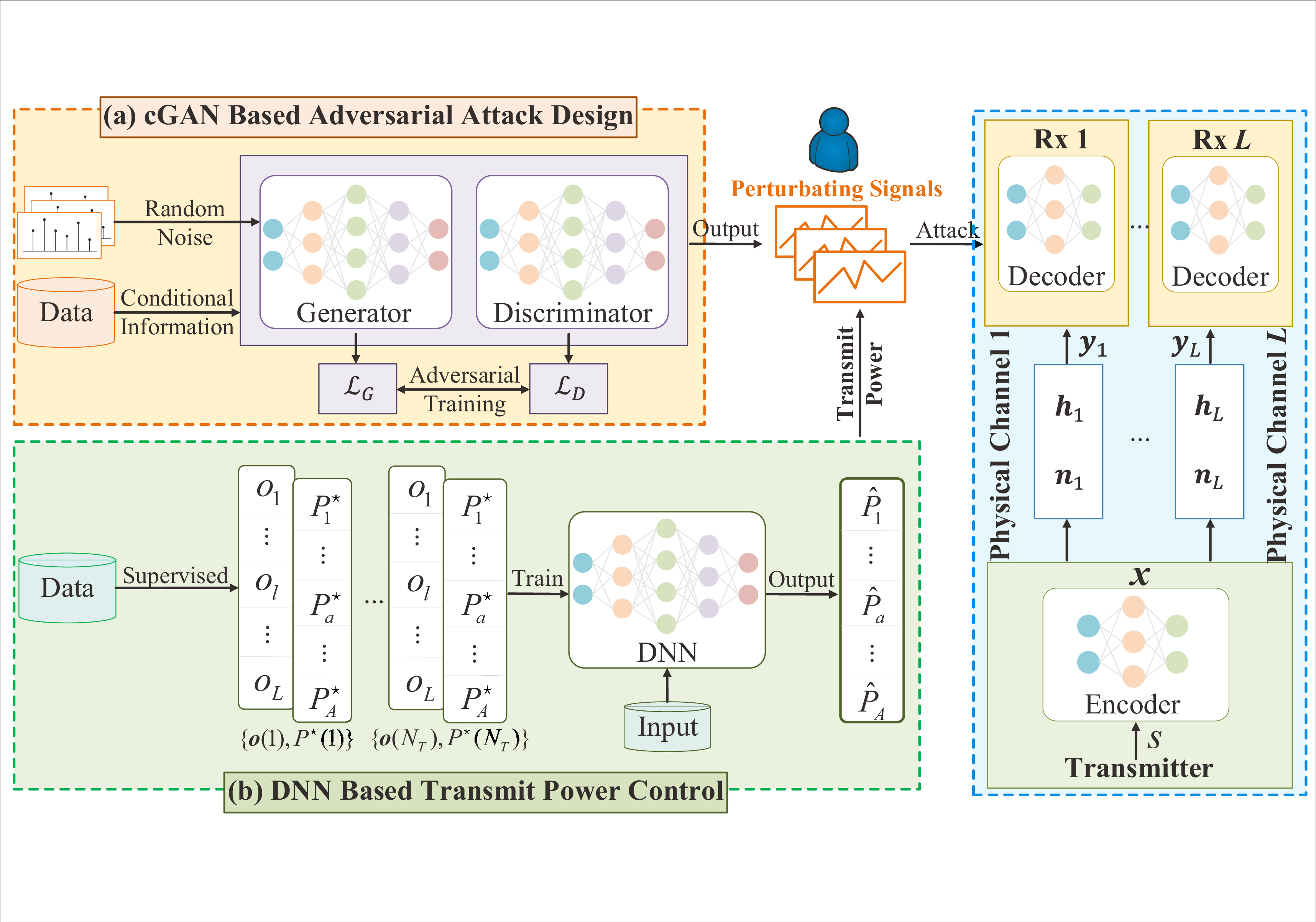}
	\caption{Overview of the DL-based framework.}
	\label{framework}
\end{figure*}
We focus on the power control to mitigate the CLI at each TR, as this term can be used to detect the attacks and the larger it is, the more likely it can be detected. Specifically, the power control problem is formulated, and the limitation of conventional optimization method is analyzed for the real-time implementation. To overcome this, a DL-based scheme is developed.

\subsection{Problem Formulation}
In the multi-adversary system, the $l$-th TR is affected not only by the direct attack from the $a$-th adversary but also by the interference leakage from other adversaries. 
Although the interference leakage introduced by other adversaries is generally weaker than the direct attack implemented by the $a$-th adversary, the aggregate leakage caused by all other adversaries cannot be neglected, leading to CLI. Such CLI can potentially alert the $l$-th TR to the existence of adversaries, thereby degrading the overall undetectability and effectiveness of attacks.
 
Accordingly, we formulate the power control problem as maximizing the product of  perturbating-signal-to-interference-plus-noise ratio (PSINR) of all TRs, given by 
\begin{equation}\label{power_control}
\begin{aligned}
\max_{P_{a,b} } \quad & \prod_{l=1}^{L}  \gamma_{l,b}^a, \\
s.t. \quad & 0 \leqslant P_{a,b} \leqslant P_{\max}.
\end{aligned}
\end{equation}
In \eqref{power_control}, $\gamma_{l,b}^a$ denotes the PSINR of the $l$-th TR attacked by the $a$-th adversary in the $b$-th time slot from \eqref{channel_output_attack} as
\begin{equation}\label{SINR}
\gamma _{l,b}^a = \frac{{{P_{a,b}}e_{l,b}^a}}{{\sum\limits_{j=1, j \neq a}^A {{P_{j,b}}r_{l,b}^j + {\sigma ^2}} }},
\end{equation}
where $\sigma^2$ is the variance of AWGN, and $e_{l,b}^a$ and $r_{l,b}^j$ represent the average attack channel gain and average interference gain, respectively, given by 
\begin{equation}\label{e}
e_{l,b}^a = \, \vert \mathbb{E} [ (\boldsymbol{h}_{l,b}^a)^{H} \boldsymbol{w}_{a,b} ] \vert^2,
\end{equation}
and 
\begin{equation}\label{r}
r_{l,b}^j = \mathbb{E} [\vert (\boldsymbol{h}_{l,b}^j)^{ H} \boldsymbol{w}_{j,b}  \vert^2], j=1,j \neq a. 
\end{equation}

\subsection{Conventional Optimization Method}
According to \eqref{SINR}, the problem \eqref{power_control} is coupled and non-convex, making it difficult to derive a closed-form solution. 
To address this issue, we introduce a positive auxiliary variable $\zeta_{l,b}^a$ and equivalently reformulate the original fractional problem into a geometric programming (GP) form as
\begin{equation}\label{power_control_DNN}
\begin{aligned}
\max_{P_{a,b},\zeta_{l,b}^a } \quad & \prod_{l=1}^{L}  \zeta_{l,b}^a, \\
s.t. \quad &  \sum\limits_{j=1, j \neq a}^A\frac{\zeta_{l,b}^a{ {{P_{j,b}}r_{l,b}^j } }}{{{P_{a,b}}e_{l,b}^a}}+\frac{\zeta_{l,b}^a \sigma ^2}{{{P_{a,b}}e_{l,b}^a}} \leqslant 1, \\
\quad &0 \leqslant P_{a,b} \leqslant P_{\max}, 
\end{aligned}
\end{equation}
which can be solved to obtain the solution to \eqref{power_control} \cite{Massive}. It is worth noting that solving \eqref{power_control_DNN} requires $r_{l,b}^j$ and $e_{l,b}^a$, depending on the following statistics and parameters:

1) Random position of TRs: $\{o_{l,b}\}_{l=1,b=1}^{L,B}$;

2) Large-scale fading coefficient: $\{\beta_{l,b}^a\}_{a=1,l=1,b=1}^{A,L,B}$;

3) Spatial correlation matrix: $\{\boldsymbol{R}_{l,b}^a\}_{a=1,l=1,b=1}^{A,L,B}$;

4) Attack channel vector estimated via MMSE: $\{\boldsymbol{\hat h}_{l,b}^{a}\}_{a=1,l=1,b=1}^{A,L,B}$;

5) Beamforming vector designed from $\{\boldsymbol{\hat h}_{l,b}^{a}\}_{a=1,l=1,b=1}^{A,L,B}$ using MMSE combining: $\{\boldsymbol{w}_{a,b}\}_{a=1,b=1}^{A,B}$;

6) Monte Carlo averaging over the estimated attack channel to compute $\{r_{l,b}^j\}_{j=1,l=1,b=1}^{A,L,B}$ and $\{e_{l,b}^a\}_{a=1,l=1,b=1}^{A,L,B}$.

As a result, whenever the TR position changes, the corresponding large-scale fading coefficient  $\beta_{l,b}^a$ and spatial correlation matrix $\boldsymbol{R}_{l,b}^a$ should be updated. This further requires reestimating the attack channel vector $\boldsymbol{\hat h}_{l,b}^{a}$ and redesigning the beamforming vector $\boldsymbol{w}_{a,b}$, after which $r_{l,b}^j$ and $e_{l,b}^a$ need to be recomputed via Monte Carlo averaging. 
Although \eqref{power_control_DNN} can be solved in polynomial time, repeatedly recomputing the required statistics and resolving the problem whenever the TR positions change is still computationally burdensome for real-time applications \cite{Downlink}.

Nevertheless, by leveraging a DNN based method, the transmit power can be obtained in real time by simply feeding the updated TR positions into the DNN, without the need to recompute $r_{l,b}^j$ and $e_{l,b}^a$ and resolve \eqref{power_control_DNN}. 
Furthermore, according to \eqref{channel_output_attack}, enhancing the attack effectiveness requires not only suppressing the CLI $\sum_{j=1, j \neq a}^{A} \sqrt{P_{j,b}}(\boldsymbol{h}_{l,b}^j)^{\rm H} \boldsymbol{w}_{j,b} \vartheta_{l,b}^j$  but also carefully designing the perturbating signal $\vartheta_{l,b}^a$, thereby reducing the risk of detection. Motivated by the above, we establish a DL-based framework shown in Fig. \ref{framework}, which comprises two main components: 1) DNN based power control, and 2) cGAN based adversarial attack design. The DNN is leveraged to estimate the transmit power of adversaries under the complex adversarial environment to enhance the undetectability, while the cGAN is utilized to carefully craft the perturbating signal to improve both the imperceptibility and aggressivity of attacks.

\subsection{DNN Based Power Control}
To reduce the high computational burden caused by repeating \eqref{power_control_DNN} in real time, a DNN based scheme is established for the real-time power control, as shown in Fig. \ref{framework}(b). The DNN is trained to learn the unknown mapping from the TR positions to the power solution of \eqref{power_control}, which is obtained through solving \eqref{power_control_DNN} via CVX. 

The proposed DNN is implemented as a feedforward neural network with fully connected layers, comprising a 2$AK$-dimension input layer, $O$ hidden layers and a $A$-dimension output layer, where $K$ represents the number of TRs attacked by each adversary and $O$ denotes the number of hidden layers. After training, the output layer of DNN can produce an estimated vector $\boldsymbol{\hat{P}} = [\hat P_1,...,\hat P_a,...,\hat P_A] \in \mathbb{R}^{A \times 1}$, which approximates the optimal solution vector $\boldsymbol{P}^{\star} = [P_1^{\star},...,P_a^{\star},...,P_A^{\star}] \in \mathbb{R}^{A \times 1}$, enabling the power estimation even when the TR positions are not included in the training phase, as illustrated in Fig. \ref{framework}(b). The training process is guided by the following loss function, defined as 
\begin{equation}\label{Loss}
\mathcal{L}(\boldsymbol{\hat P}(n), \boldsymbol{P}^{\star}(n)) = \frac{1}{A}\sum_{i=1}^{A} \left(\frac{\boldsymbol{\hat P}_i (n) - \boldsymbol{P}^{\star}_i (n)}{\boldsymbol{P}^{\star}_i (n)}\right)^2,
\end{equation}
where $\boldsymbol{\hat P}_i (n)$ and $\boldsymbol{P}^{\star}_i (n)$ represent the $i$-th components of the predicted and optimal transmit power vectors at the $n$-th sample, respectively, $n \in \{1,...,N_T\}$, with $ N_T \in \mathbb{Z}^+$ denoting the size of training dataset.  

Accordingly, the training objective is to minimize the average loss over the entire training dataset as
\begin{equation}\label{train_objective}
\min_{\boldsymbol{W},\boldsymbol{b}} \frac{1}{N_T} \sum_{n=1}^{N_T} \mathcal{L}(\boldsymbol{\hat P}(n), \boldsymbol{P}^{\star}(n)),
\end{equation}
where $\boldsymbol{W}$ and $\boldsymbol{b}$ denote the weights and biases of DNN, respectively. As a result, the power control problem can be reformulated as a parameter learning task, in which $\boldsymbol{W}$ and $\boldsymbol{b}$ of DNN are optimized to approximate the input-output mapping obtained from solving \eqref{power_control_DNN}. Once $\boldsymbol{W}$ and $\boldsymbol{b}$ have been configured, the trained DNN can directly estimate the transmit power based on the TR positions without the need to resolve \eqref{power_control_DNN}.

\textit{Remark:} During the offline labeling and training, a centralized adversarial controller is assumed to have access to the global information required for the power optimization, including the TR positions, estimated attack channel vectors, beamforming vectors, noise variance and power threshold. This information is utilized to compute $r_{l,b}^j$ and $e_{l,b}^a$ and solve \eqref{power_control_DNN} to obtain the position-power labeled dataset, which is subsequently used to train the DNN to learn the mapping from the TR positions to the optimal transmit power solution. During the online inference, the trained DNN parameters remain fixed, and only the TR positions are required as the input to DNN. By separating the offline labeling and training from the online inference, the proposed DNN-based scheme avoids repeated statistical computation and GP solving, facilitating the real-time transmit power control in the dynamic environment.



\section{cGAN Based Adversarial Attack}
In this section, we first formulate the adversarial attack problem, and provide a brief review on cGAN. Then, a cGAN based adversarial attack scheme is proposed to generate the perturbating signal that not only adapts to the dynamic attack channel but also introduces only a small deviation from the original waveform.

\subsection{Problem Formulation}
By introducing an adversarial perturbating signal, the decoder at the TR may make a wrong estimation as
\begin{equation}
\mathcal{G}(\boldsymbol{y}_l) \neq \mathcal{G}(\boldsymbol{y}_l^a),
\end{equation} 
which demonstrates the effectiveness of adversarial attacks. 

To improve the attack effectiveness, the amplitude of perturbating signal $\vartheta_{l,b}^a$ is required to be large to induce the misclassification. On the other hand, its amplitude should also remain small to be imperceptible at the TR and enable the undetectability. Therefore, the adversarial perturbating signal $\boldsymbol{\vartheta}_l^a$ from the $a$-th adversary to the $l$-th TR during the transmission should be well designed, and the problem can be formulated as
\begin{equation}\label{p0}
\begin{aligned}
\min_{\vartheta_{l,b}^a } \quad & \|\boldsymbol{\vartheta}_l^a\|_2 \\
s.t. \quad & \mathcal{G}(\boldsymbol{y}_l) \neq \mathcal{G}(\boldsymbol{y}_l^a).	
\end{aligned}
\end{equation}

Although the objective in \eqref{p0} is convex, the misclassification constraint in \eqref{p0} corresponds to a discrete 0-1 condition, which is non-differentiable and leads to a non-convex feasible set. Moreover, since $\mathcal{G}(\cdot)$ is implemented via neural networks, its nonlinear decision boundary further exacerbates the non-convexity. Consequently, the problem \eqref{p0} does not have a general closed-form solution and cannot be solved via standard convex optimization methods. In addition, the design of the perturbating signal $\vartheta_{l,b}^a$ should consider both the attack effectiveness and undetectability. Specifically, a perturbating signal with higher amplitude is more likely to induce misclassification, while a smaller amplitude helps preserve the waveform similarity and reduce the risk of detection. This inherent tradeoff becomes more challenging under the dynamic adversarial environment, where the variations in the attack channel may alter the effectiveness of a given perturbating signal, making it difficult to maintain consistent attack performance. To address these challenges, a cGAN scheme is established, in which the generation process is conditioned on the attack channel information. This enables the generated perturbating signal to adapt to the attack channel variation while taking into account both the attack effectiveness and undetectability.
\begin{figure}[t]
	\centering
	\includegraphics[width=3.4in]{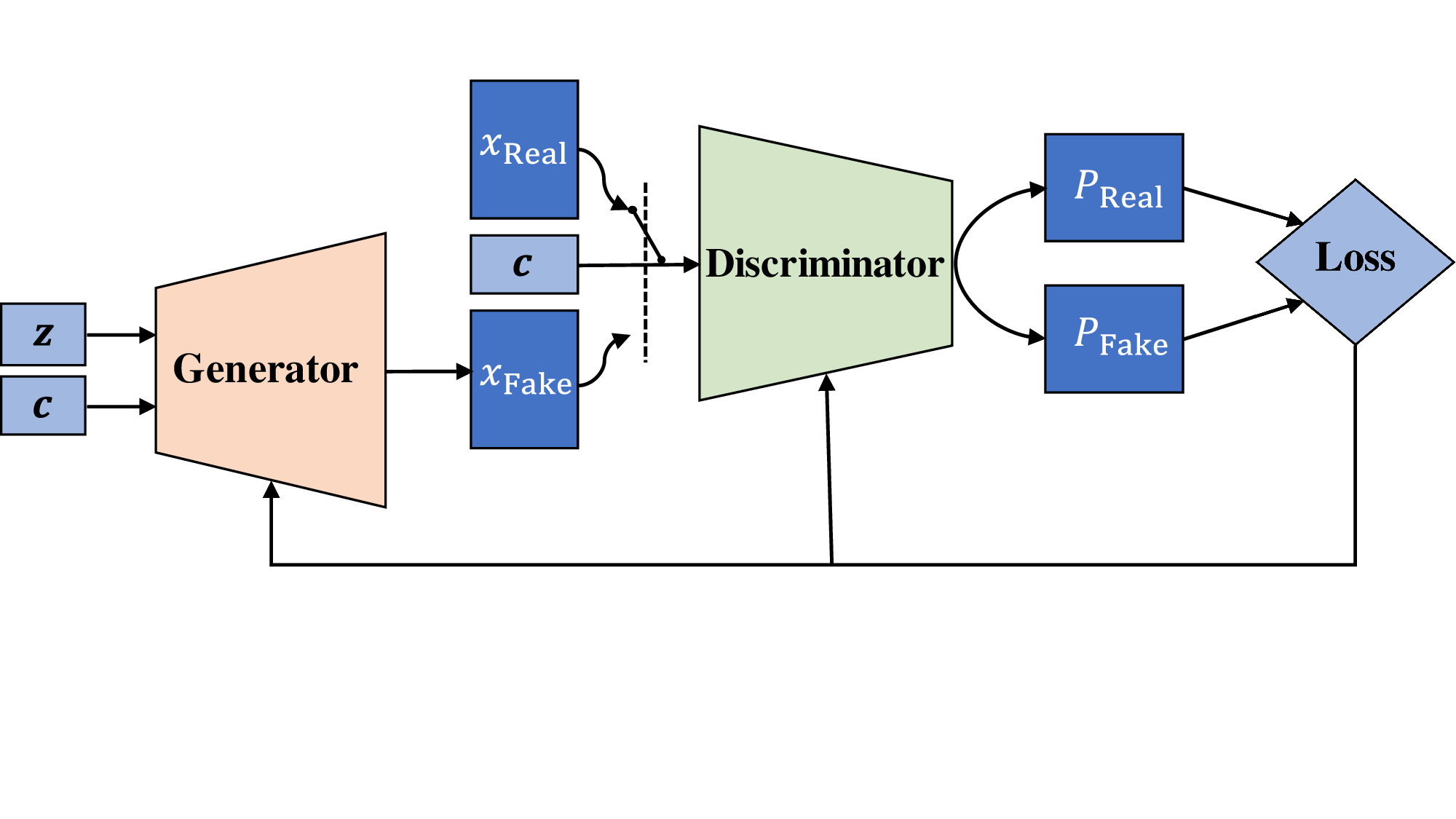}
	\vspace{-41pt}
	\caption{Framework of cGAN.}
	\label{CGAN}
\end{figure}
\begin{figure*}[htbp]
	\centering
	\includegraphics[width=6.5in]{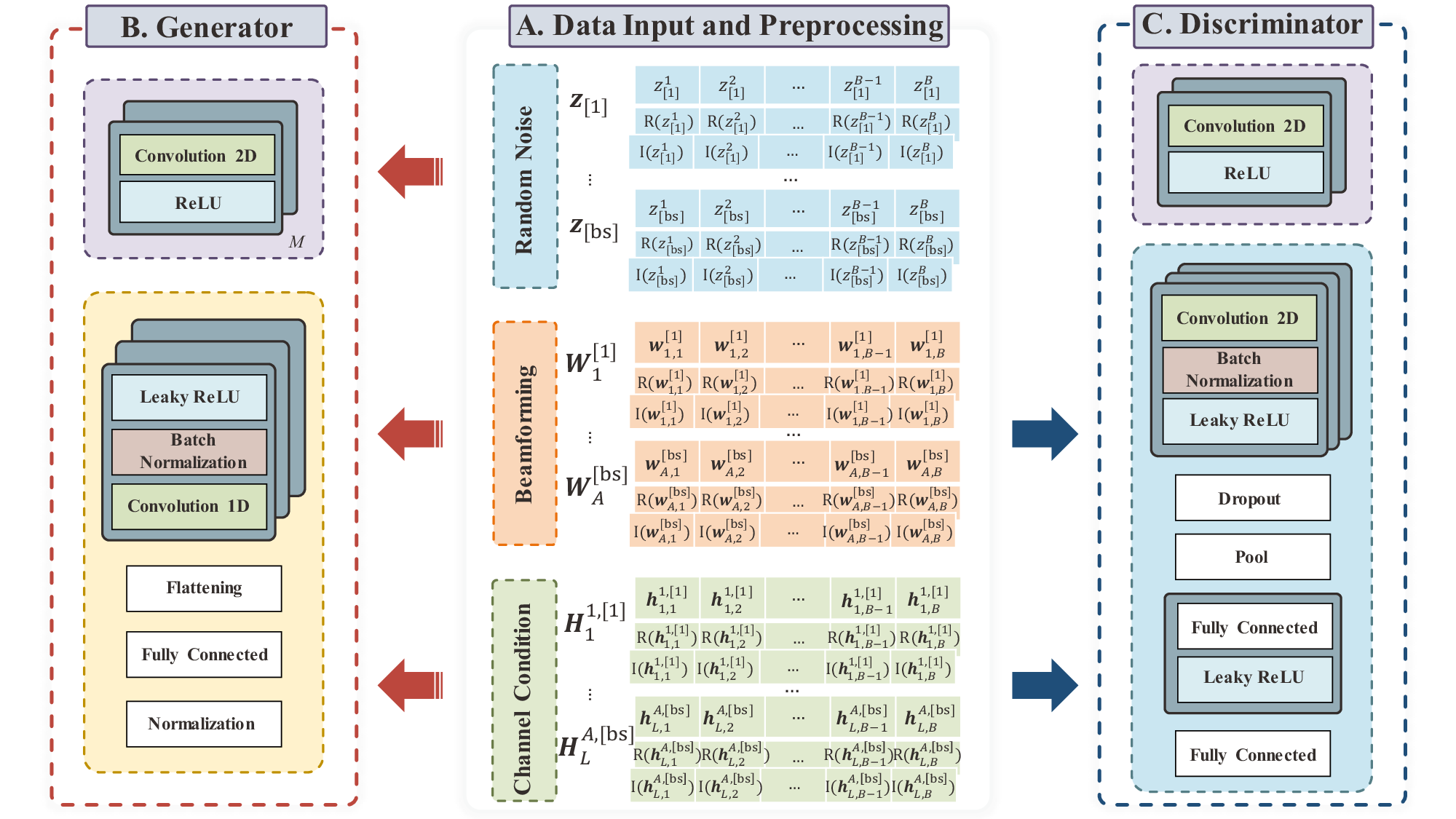}
	\caption{Network architectures of data preprocessing, generator, and discriminator.}
	\label{generator}
\end{figure*}

\subsection{Brief Review on cGAN}

cGAN is an extension of the original generative adversarial network (GAN) to introduce the conditional information to generate fake samples similar to the real ones, and has been widely utilized to generate various digital contents \cite{Zhang}. As shown in Fig. \ref{CGAN}, the cGAN consists of two neural networks: a generator $G$ and a discriminator $D$. The generator produces fake samples to resemble the real ones from the random noise and corresponding conditional information. In contrast, the discriminator, as a classifier, focuses on distinguishing between the real samples and fake ones created by the generator.

More specifically, the generator aims to maximize the probability that generated samples are classified as the real ones by the discriminator. The discriminator has two tasks: 1) maximize the probability of correctly classifying the real samples as real ones, and 2) minimize the probability of classifying the generated samples as real ones.

Accordingly, the objectives of generator and discriminator can be respectively formulated as
\begin{equation}\label{G}
\min_G \  \mathbb{E}_{\boldsymbol z \sim f}  \log(1 - D(G(\boldsymbol z|\boldsymbol c))),
\end{equation}
\begin{equation}\label{D}
\max_D \ \mathbb{E}_{\boldsymbol x \sim \tilde{f}}  \log D(x)  + \mathbb{E}_{\boldsymbol z \sim f}  \log(1 - D(G(\boldsymbol z|\boldsymbol c))),
\end{equation}
where $f$ and $\tilde{f}$ denote the distribution of generative model and real samples, respectively, $\boldsymbol x$, $\boldsymbol z$ and $\boldsymbol c$ represent the real sample, random noise and conditional information, respectively, and $G(\boldsymbol z|\boldsymbol c)$ is the generated fake samples.

In the training period, the generator and discriminator are alternately optimized in an adversarial manner as
\begin{align}
\min_G \max_D \ V(D, G) 
&= \min_G \max_D [ \mathbb{E}_{\boldsymbol x \sim \tilde{f}} \log D(x) \nonumber \\
&\quad + \mathbb{E}_{\boldsymbol z \sim f} \log\!\left(1 - D(G(\boldsymbol z|\boldsymbol c))\right) ],
\end{align}which can enhance the performance of both the generator and discriminator. Finally, the generator and discriminator can reach the Nash Equilibrium, when the distribution of fake samples will be close to that of real ones.

\subsection{cGAN Based Adversarial Attack Design}
The proposed cGAN based adversarial attack scheme is shown in Fig. \ref{framework}(a), and the detailed framework of cGAN is illustrated in Fig.~\ref{CGAN}, which comprises four main components: data input and preprocessing, generator, discriminator and network training.

\textit{1) Data Input and Preprocessing:}
The input data consist of two components: the random noise $\boldsymbol{z}$ and the conditional information $\boldsymbol{c}$. The latter contains the beamforming and attack channel vectors. Since the random noise, beamforming and attack channel vectors are complex-valued, which are difficult for cGAN to process directly, they are decomposed into real and imaginary parts before being fed into the cGAN, as shown in Fig. \ref{generator}A.
	
\textit{2) Generator:} 
As shown in Fig. \ref{generator}B, the generator comprises two functional modules. In the first module, the conditional information $\boldsymbol{c}$, i.e., the beamforming and attack channel vectors, is processed to extract the feature, which can characterize the attack channel. The second module integrates the random noise $\boldsymbol{z}$ and the output of first module, refines the joint feature through an enhanced convolutional network, and maps it into a target dimension via the flattening and dense layers, ultimately producing the normalized perturbating signal.

\textit{3) Discriminator:} The discriminator aims to distinguish between the clean and attacked received signals, with its architecture depicted in Fig. \ref{generator}C. Specifically, it takes the same conditional information $\boldsymbol{c}$ as the generator to maintain the conditional consistency, and extracts the feature from both the beamforming and attack channel vectors. Then, the resulting conditional representation is concatenated with the attacked received signal, and fed into a stack of convolutional layers with progressively increased filter sizes, each followed by the batch normalization and Leaky ReLU activation. Afterwards, a dropout layer is introduced to mitigate the overfitting, and the global average pooling is utilized to aggregate the output of dropout layer into a compact representation. Finally, two fully connected layers are employed. The first is equipped with the Leaky ReLU activation and produces a 256-dimension feature representation, while the second outputs a final classification result with a dimension of $2^k$. Notably, the clean received signal is processed through the same discriminator as the attacked received signal, thus enabling the comparability between them.

\textit{4) Network Training:}  According to \eqref{G} and \eqref{D}, the loss functions of discriminator and generator can be reformulated as
\begin{align}
\mathcal{L}_D =\ 
& \mathbb{E}_{\boldsymbol{x} \sim \tilde{f}} [ \log D({\boldsymbol{y}}_l|\boldsymbol c) ] + \mathbb{E}_{\boldsymbol{z} \sim f} [ \log (1 - D( {\boldsymbol{y}}_l^a|\boldsymbol c ) ) ],
\label{D_loss}
\end{align}	
\begin{equation}
	\mathcal{L}_G = \mathbb{E}_{\boldsymbol z \sim f}  \log(1 - D({\boldsymbol{y}}_l^a |\boldsymbol c)).
\end{equation}

The discriminator and generator are trained in an alternating manner. Specifically, the discriminator is updated by maximizing $\mathcal{L}_D$ to enhance its ability to distinguish the clean received signal from the attacked one under the given condition $\boldsymbol c$, and the generator is updated by minimizing $\mathcal{L}_G$ to produce the perturbating signal that can mislead the discriminator. Guided by the conditional information $\boldsymbol c$, the generator adapts the generated perturbating signal to the dynamic attack environment. Through this adversarial training process, the generator progressively learns to produce the perturbating signal that can induce the misclassification while remaining difficult to be detected.

\textit{5) Evaluation Metrics:}
The attack success rate (ASR) is introduced to evaluate the aggressivity of adversarial attacks. 
It is defined as the proportion of successful attack samples among all attack samples, where an attack is considered successful only when it induces the decoding failure while satisfying the power control.
In particular, the power control is regarded as successful when the estimated transmit power does not exceed the threshold during the transmission. To describe this, we introduce an indicator $\chi_a(n)$ as
\begin{equation}
\chi_a(n)  = \xi\!\left( \hat P_{a,b}(n)  \leq P_{a,b}^{\star}(n) ,\ \forall b \right),
\end{equation}
where $\chi_a(n) =1$ indicates that the power control is successful during the transmission for the $n$-th attack sample, and $\chi_a(n) =0$ otherwise, and $P_{a,b}^{\star}(n)$ is the threshold.

Accordingly, the ASR can be formulated as
\begin{equation}\label{attackrate}
\text{ASR} = \frac{1}{S_m} \sum_{n=1}^{S_m}
\xi\!\left( g(\boldsymbol{y}_l^{a}(n)) \neq s(n) \right)\chi_a(n),
\end{equation}
where $S_m$ denotes the total number of attack samples, $\boldsymbol{y}_l^{a}(n)$ denotes the attacked received signal in the $n$-th sample, $s(n)$ is the corresponding original message, and $\xi(\cdot)$ is the indicator function that equals 1 when the specified condition is satisfied and 0 otherwise.

\textit{Remark:} Rather than explicitly solving the non-convex problem \eqref{p0}, the proposed cGAN based scheme provides an implicit solution by creating the adversarial perturbating signals conditioned on the attack channel information. Specifically, the adversarial training mechanism of cGAN enables the generator to create the perturbating signals balancing the attack effectiveness and undetectability, while the conditional input guides the perturbating signals to adapt to the attack channel variations. As a result, the proposed scheme can produce the perturbating signals with strong aggressivity and low detectability in the dynamic adversarial environment.

\section{Numerical Results and Discussion}
\begin{figure*}[htbp] 
	\centering
	\begin{minipage}{0.47\textwidth} 
		\centering
		\includegraphics[width=\textwidth]{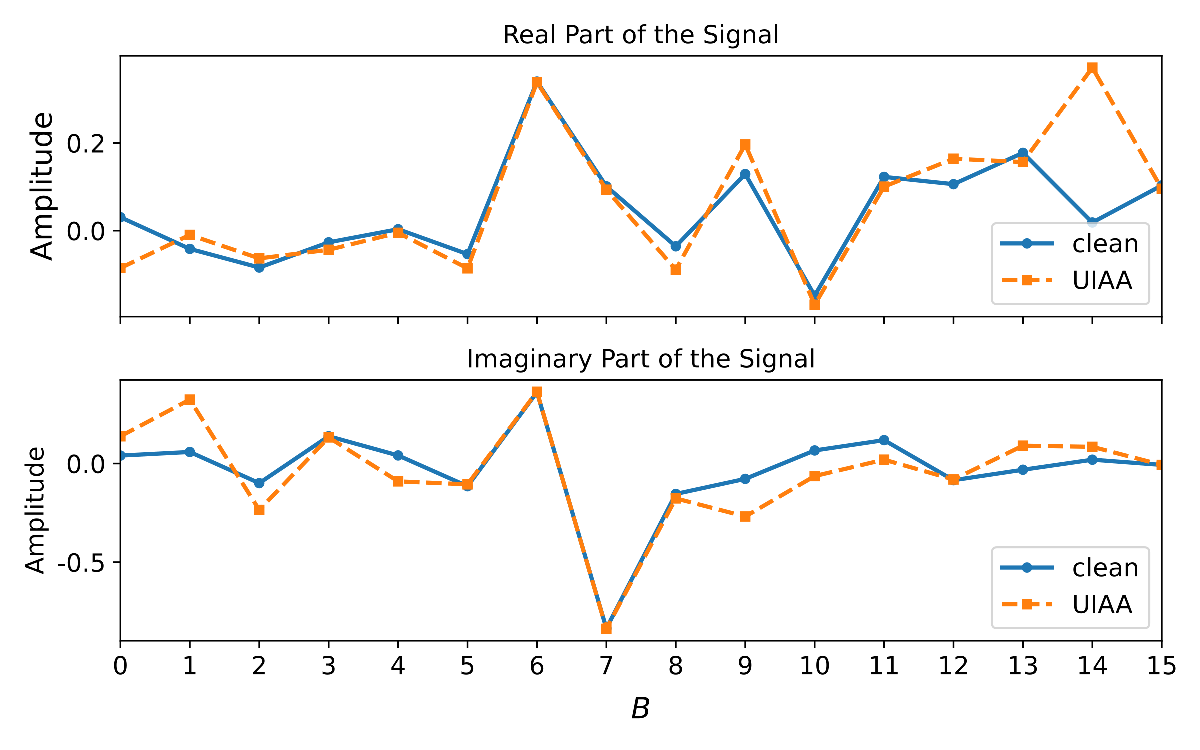} 
		\small\text{(a)}
		\includegraphics[width=\textwidth]{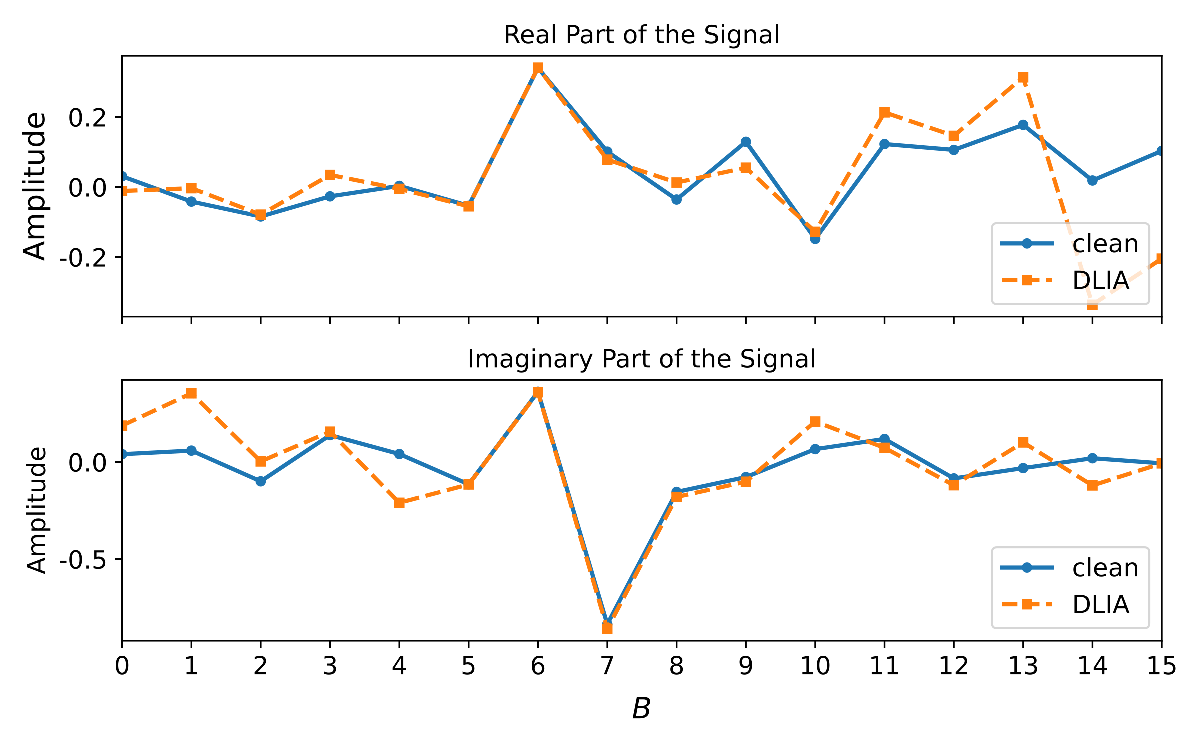} 
		\small\text{(b)}
		\includegraphics[width=\textwidth]{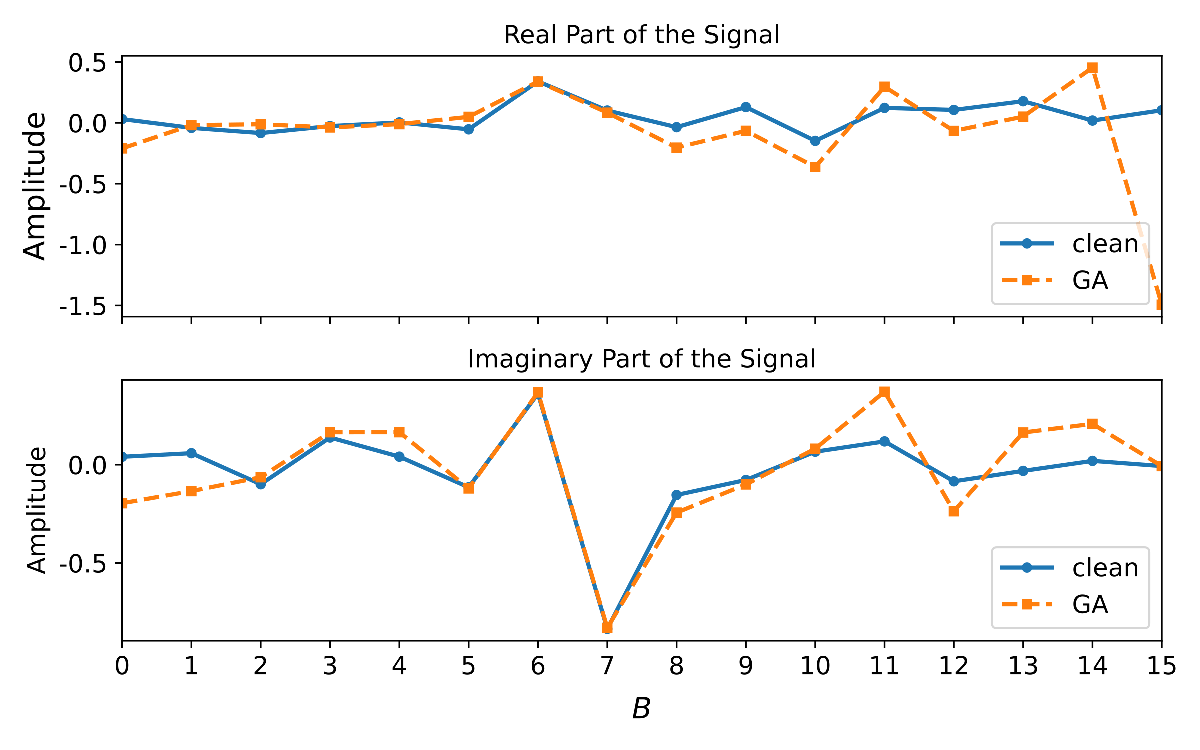} 
		\small\text{(c)}
	\end{minipage}%
	\hfill
	\begin{minipage}{0.47\textwidth} 
		\centering
		\includegraphics[width=\textwidth]{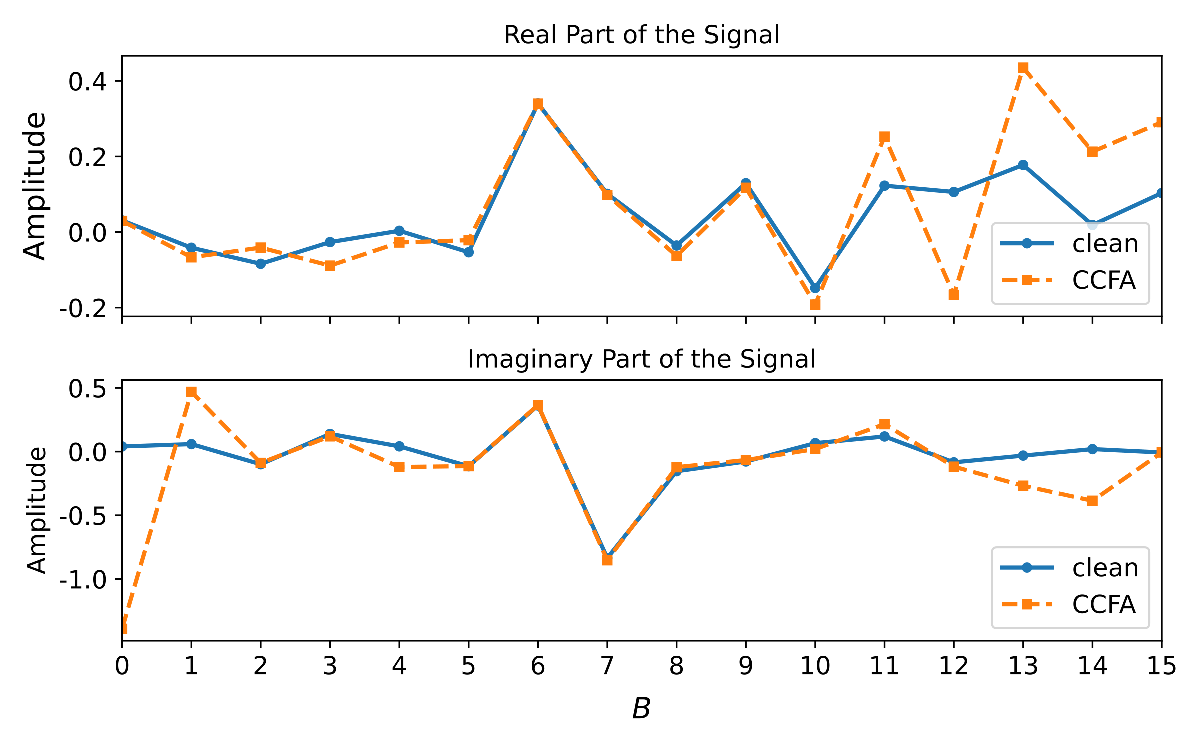} 
		\small\text{(d)}
		\includegraphics[width=\textwidth]{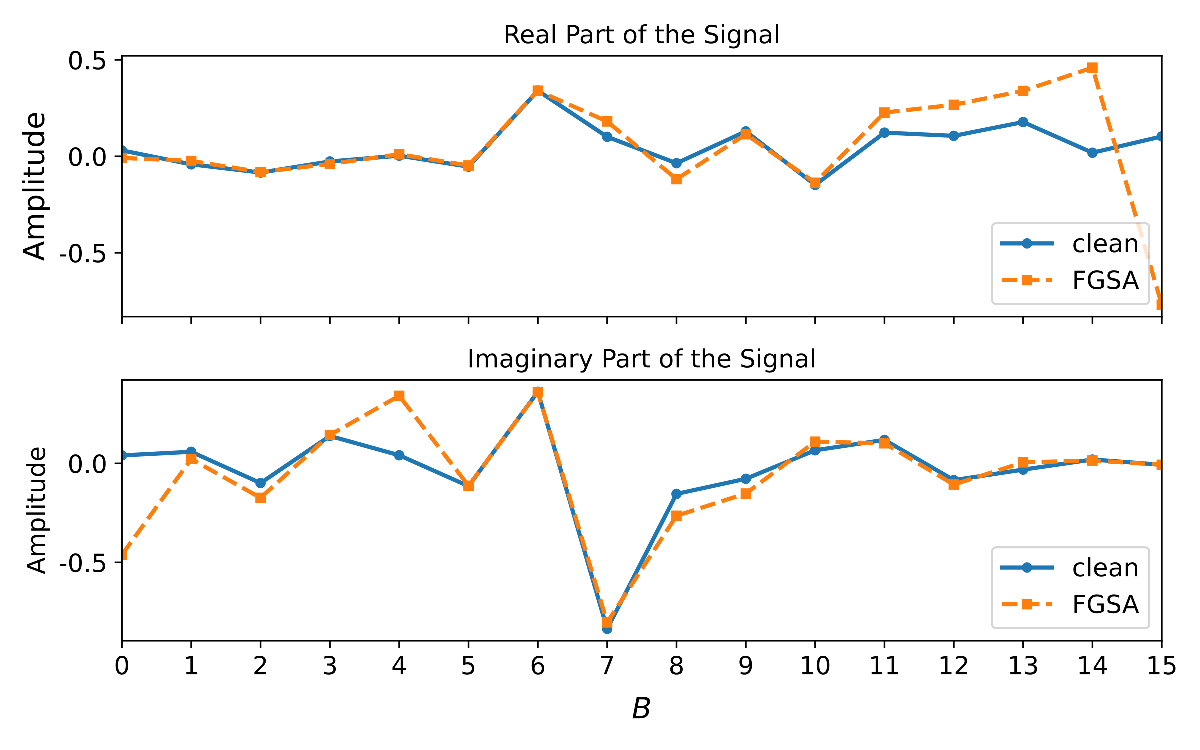}
		\small\text{(e)}
		\includegraphics[width=\textwidth]{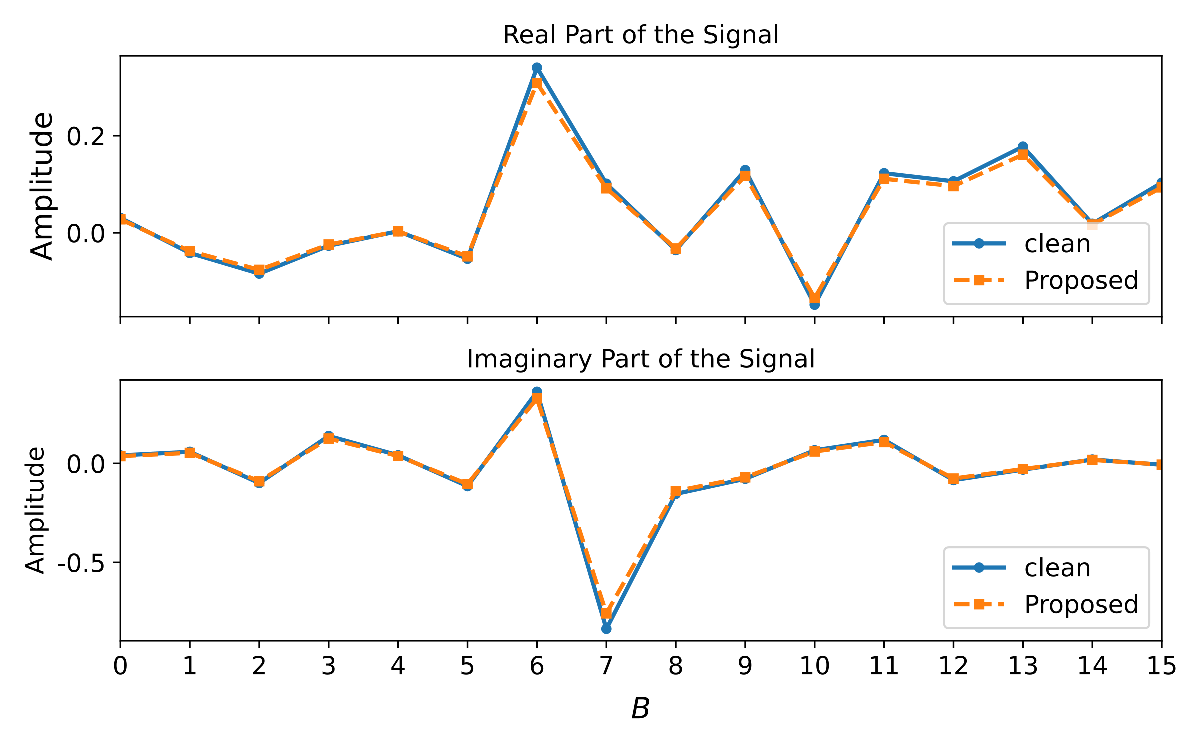} 
		\small\text{(f)}
	\end{minipage}
	\caption{Waveforms of the clean received signal and corresponding attacked received signal derived from different attack methods, including (a) UIAA, (b) DLIA, (c) GA, (d) CCFA, (e) FGSA and (f) the proposed method.}
	\label{waveform}
\end{figure*}

In this section, we first describe the experimental setup, including the datasets, baseline methods and parameter setting, and then verify the effectiveness of the proposed framework in terms of undetectability, aggressivity and adaptability. Finally, an ablation study is conducted and presented to validate the necessity of its key components.

\subsection{Experimental Setup}
\subsubsection{Datasets}
The DNN is trained and evaluated respectively on a training dataset $H$ and a test dataset $J$, both of which are obtained by solving \eqref{power_control_DNN} via CVX. Specifically, $H = \{\boldsymbol{o}(n),\boldsymbol{P}^{\star}(n)\}_{n=1}^{N_T}$ and $J = \{\boldsymbol{o}_t (n),\boldsymbol{P}^{\star}_t (n)\}_{n=1}^{N_t}$, where $H \cap J = \emptyset$, $N_t \in \mathbb{Z}^+$ is the size of test dataset, $\boldsymbol{o}(n)$ and $\boldsymbol{o}_t (n)$ represent the $n$-th samples of the DNN input during the training and testing, respectively, and $\boldsymbol{P}^{\star}(n)$ and $\boldsymbol{P}^{\star}_t (n)$ denote their corresponding ground-truth labels. In addition, $3.125\%$ of the training dataset is set aside as the  validation dataset to monitor the performance of DNN during training.

For the cGAN, the training, validation and test datasets are denoted as $H_c = \{\boldsymbol{Y}_{l} (n), \boldsymbol{Y}_l^a (n)\}_{n=1}^{N_H}$, $V_c = \{\boldsymbol{Y}_{l} (n), \boldsymbol{Y}_l^a (n)\}_{n=1}^{N_V}$ and $J_c = \{\boldsymbol{Y}_{l} (n), \boldsymbol{Y}_l^a (n)\}_{n=1}^{N_J}$, respectively, where $\boldsymbol{Y}_{l}=\{\boldsymbol{y}_{l}(b)\}_{b=1}^B$ and $\boldsymbol{Y}_l^a=\{\boldsymbol{y}_l^a(b)\}_{b=1}^B$ denote the clean and attacked received signals across the transmission period, respectively, $H_c \cap V_c  = \emptyset$, $ H_c \cap J_c = \emptyset$ and $ V_c \cap J_c = \emptyset$, and $n$ denotes the $n$-th sample.

\subsubsection{Baselines}
For comparison, consider five baselines, including the universal-class input-agnostic adversarial attack (UIAA), the double loop iterative attack (DLIA) \cite{Cognitive_IOT}, the fast gradient sign attack (FGSA) \cite{sadeghi2018adversarial}, the cGAN based channel-robust and frequency-selective attack (CCFA) \cite{Channel_RobustSISO6} and the GAN based attack (GA). They are briefly described as follows.

$\bullet$ UIAA: A universal-class iterative method that utilizes the bisection to determine the step size in each iteration.
 
$\bullet$ DLIA: A two-loop iterative method that introduces the momentum accumulation and utilizes both the outer and inner loop numbers to update the step size in each iteration.

$\bullet$ FGSA: A gradient-based method that attacks the input along the sign of the loss gradient with a single-step update. The step size is calculated using the bisection.

$\bullet$ CCFA: A cGAN based method that adapts to the dynamic attack channel and applies a Butterworth lowpass filter to suppress the high-frequency components.

$\bullet$ GA: A GAN based adversarial attack method.

\subsubsection{Parameter Setting}
$Q = B = 16$, $k=4$, $R=k/Q$, $L = A =4$, $M=N=16$, $O=3$, $ASD = 10^{\circ}$, $d_0 = 1 \, \rm km$, $P_{\max} = 10 \, \rm mW$,  $ S = 10^4$, $N_T = 10^5$, $N_t = 5000$, $N_H = 3500$, $N_V = 500$ and $N_J = 1000$. For fair comparison and efficient calculation, the number of iterations in UIAA is fixed at $10$, the number of momentum accumulations in DLIA is set to $10$, the number of outer and inner loops in DLIA are set to $2$ and $5$, respectively, and the number of bisection iterations in UIAA and FGSA is set to $30$. 

$\bullet$ Hyperparameters of DNN: The batch size is set to $256$, and the optimizer is Adam with an initial learning rate of $1e-2$. During the training, learning rate is gradually reduced to $1e-3$, $1e-4$ and $1e-5$. Early stopping is adopted to prevent the overfitting, with a patience of $50$ epochs.

$\bullet$ Hyperparameters of cGAN: The training parameters include batch size, learning rate and epochs. Specifically, four training configurations are adopted, as shown in Table \ref{tab:hyperparameter_settings}. Each training configuration is evaluated with the same validation configuration, where the validation batch size and validation interval are set to $128$ and $5$, respectively. These hyperparameter settings are empirically selected to improve the model performance while maintaining the computational efficiency. 

\begin{table}[t]
	\centering
	\vspace{2mm}   
	\caption{cGAN training configurations.}
	\label{tab:hyperparameter_settings}
	\setlength{\tabcolsep}{10pt}      
	\renewcommand{\arraystretch}{1.2} 
	\begin{tabular}{lll}
		\hline
		\textbf{Batch size} & \textbf{Learning rate} & \textbf{Epochs} \\
		\hline
		$64$  & $1e-3$ & $120$ \\
		$64$  & $1e-4$ & $120$ \\
		$128$ & $1e-4$ & $150$ \\
		$128$ & $1e-5$ & $150$ \\
		\hline
	\end{tabular}
\end{table}
\begin{figure}[htbp]   
	\hspace*{-0.20in}          
	\includegraphics[width=3.6in]{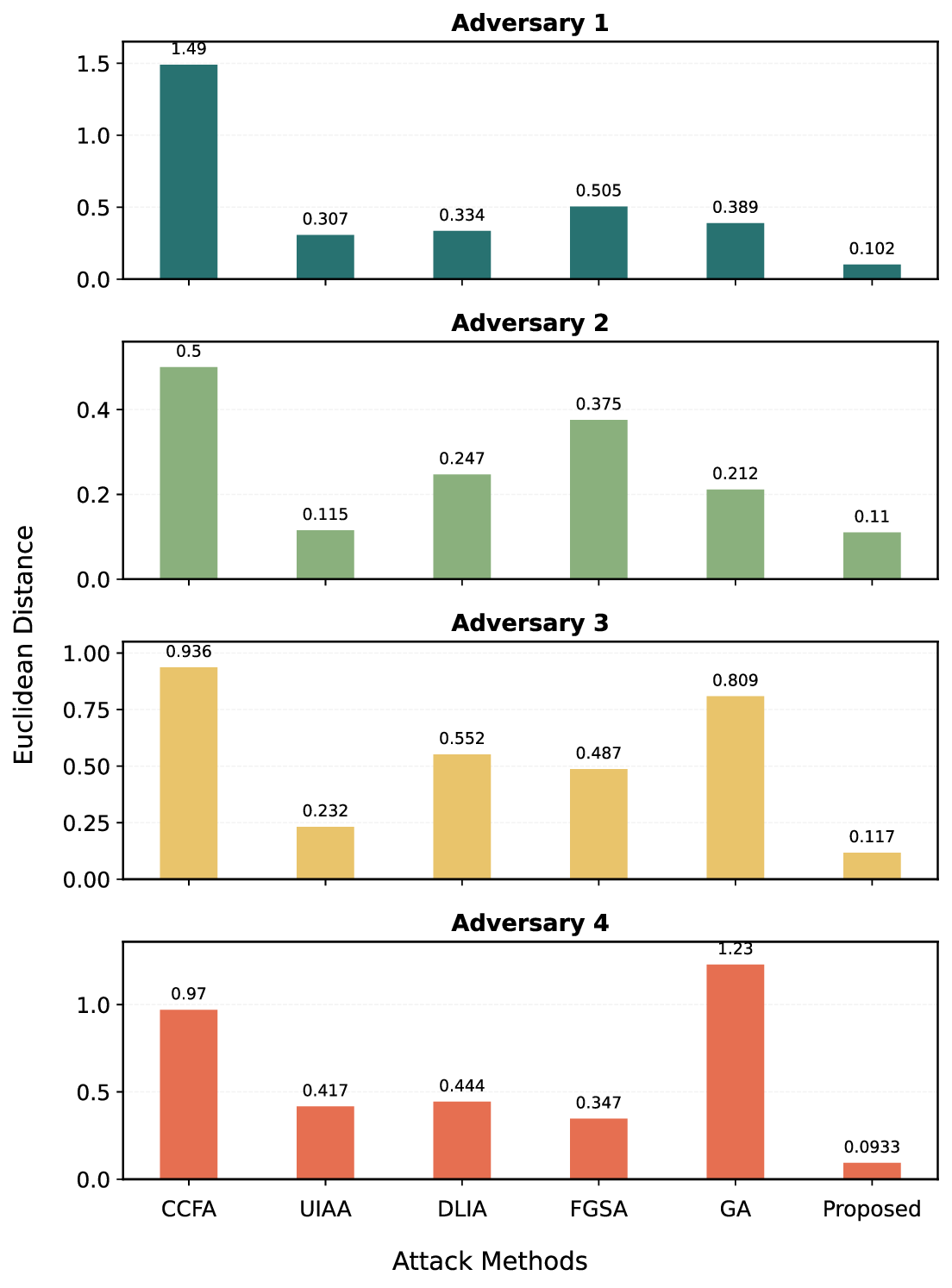}
	\caption{Euclidean distance between the clean and corresponding attacked received signals under diverse attack methods, including CCFA, UIAA, DLIA, FGSA, GA and the proposed scheme.}
	\label{euclidean_distance}
\end{figure}

\begin{figure}[htbp]     
	\centering         
	\includegraphics[width=3.5in]{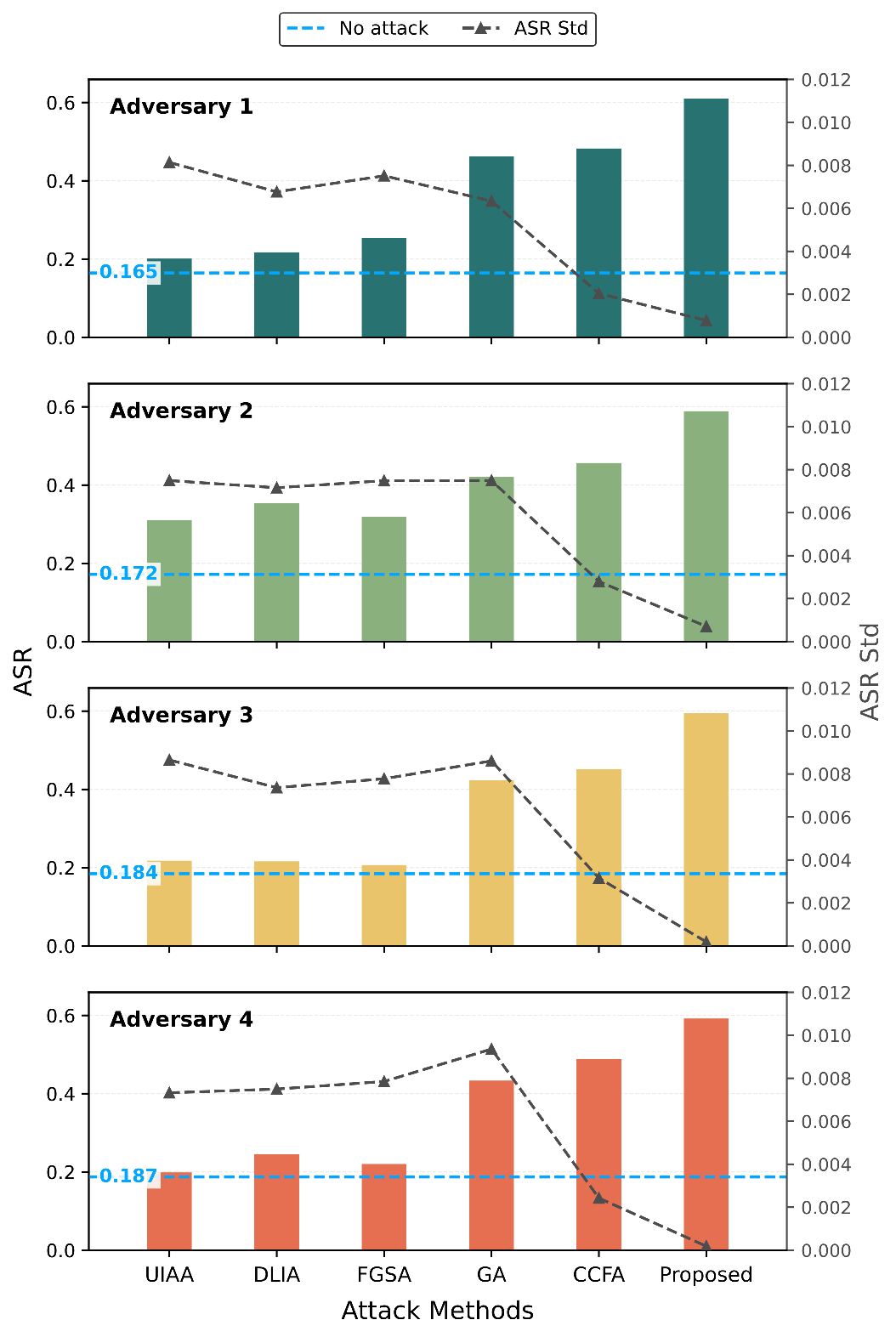}
	\caption{ASR and ASR Std under diverse attack methods, including CCFA, UIAA, DLIA, FGSA, GA and the proposed scheme.}
	\label{attack}
\end{figure}

\begin{figure}[htbp]     
	\centering         
	\includegraphics[width=3.10in]{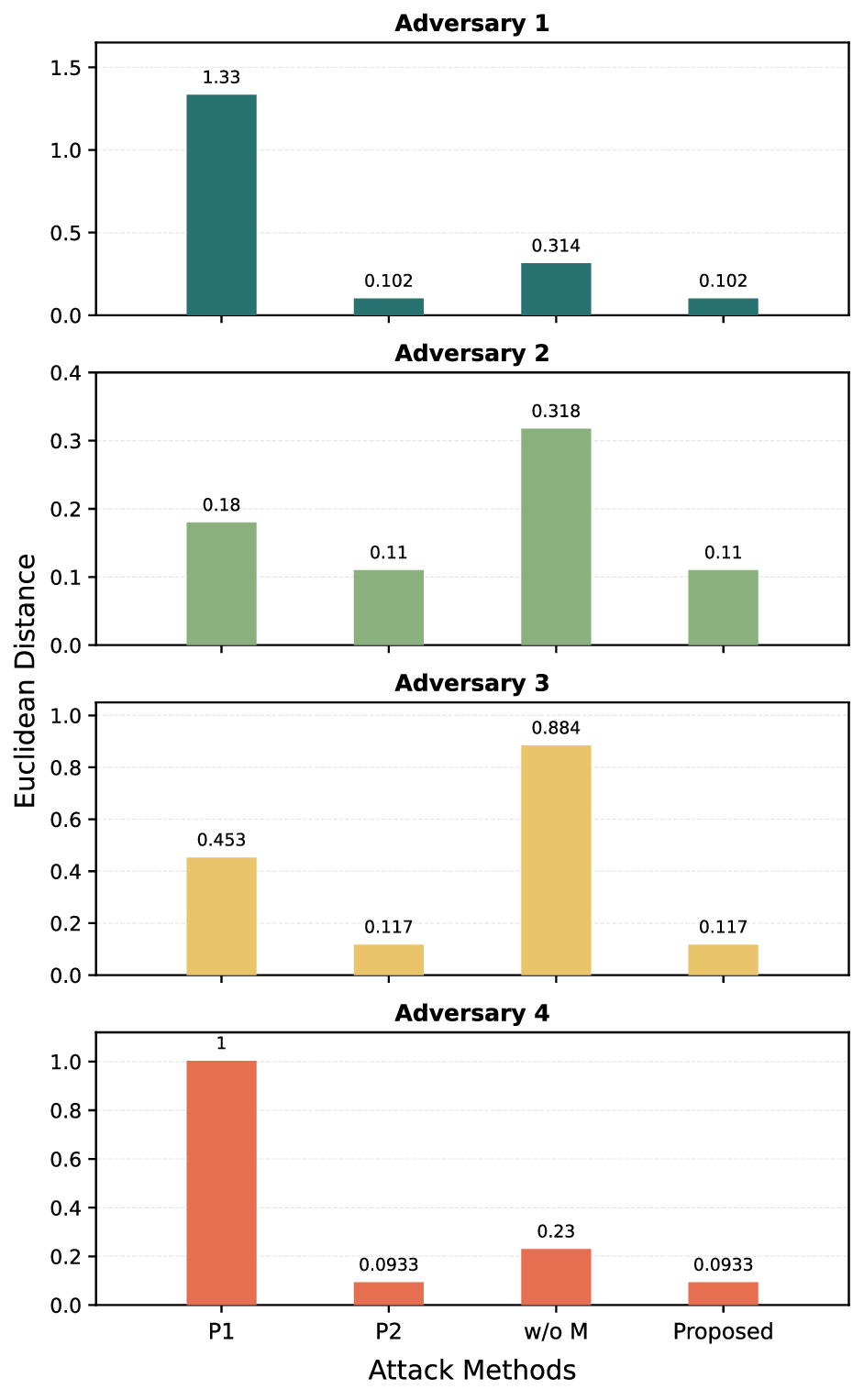}
	\caption{Euclidean distance between the clean and corresponding attacked received signals under diverse attack methods, including \textbf{P1}, \textbf{P2}, \textbf{w/o M} and the proposed scheme.}
	\label{ed_}
\end{figure}
\begin{figure}[htbp]                 
	\includegraphics[width=3.5in]{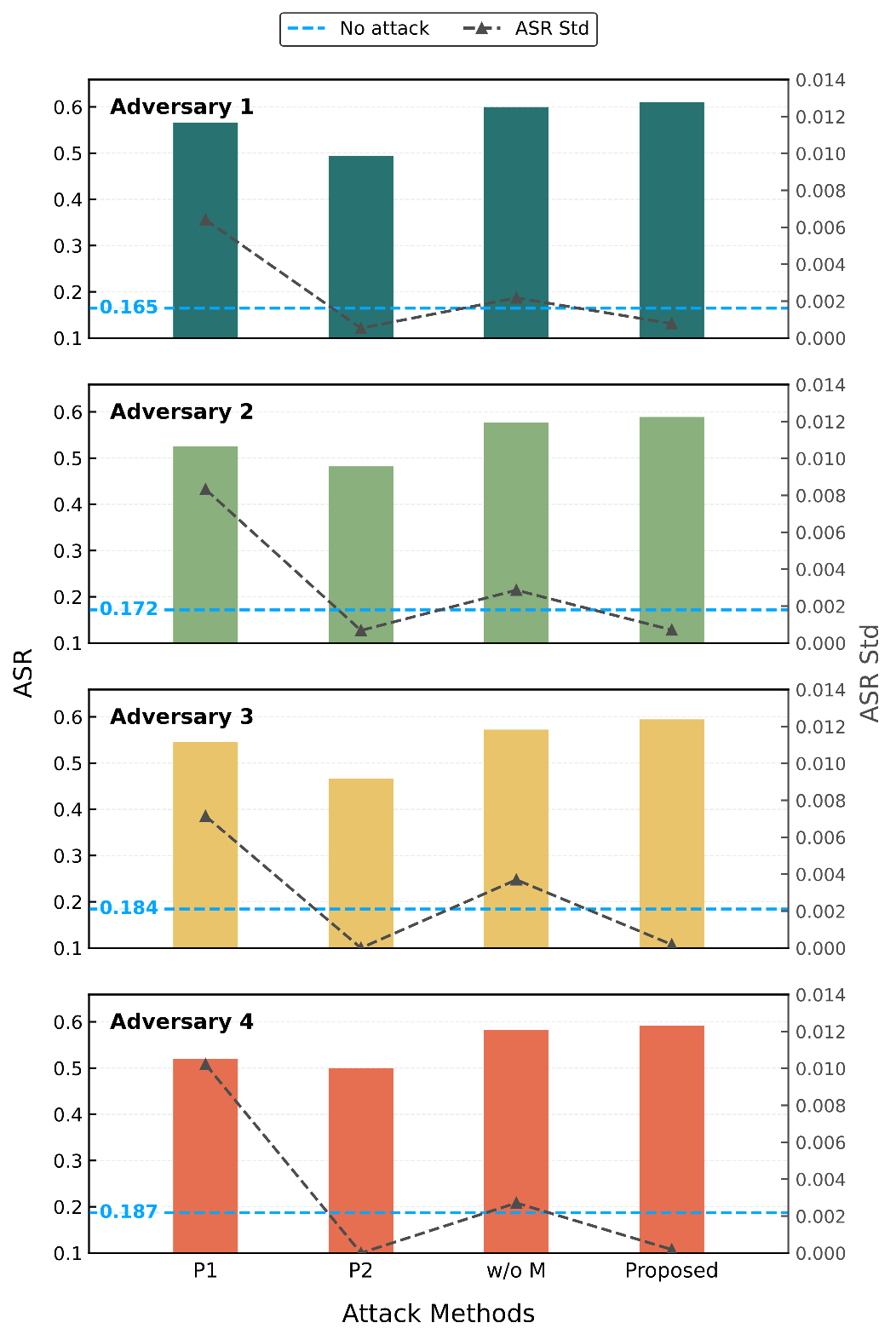}
	\caption{ASR and ASR Std under diverse attack methods, including \textbf{P1}, \textbf{P2}, \textbf{w/o M} and the proposed scheme.}
	\label{asr-part}
\end{figure}

\subsection{Undetectability Comparison}
Taking Adversary 1 as an illustrative example, Fig.~\ref{waveform} presents the waveforms of clean received signal and corresponding attacked received signal under different attack methods, including the CCFA, UIAA, DLIA, FGSA, GA and the proposed scheme. For clarity, the complex signal is decomposed into real and imaginary parts for comparison.  
As shown in Fig. \ref{waveform}, the waveform of the attacked received signal obtained by the proposed scheme is the most similar to that of the clean received signal.
To further evaluate the performance, we introduce the Euclidean distance to quantitatively measure the discrepancy between the clean and attacked received signals in Fig.~\ref{euclidean_distance}. From the results, we can see that for each adversary, the waveform of the attacked received signal derived by the proposed scheme exhibits the highest similarity to that of the clean received signal.

\subsection{Aggressivity and Adaptability Comparison}
Fig. \ref{attack} presents the ASR and ASR standard deviation (Std) of CCFA, UIAA, DLIA, FGSA, GA and the proposed scheme evaluated on the test dataset.
ASR evaluates the aggressivity of attack methods, while ASR Std reflects their adaptability to dynamic attack channels.
As shown in Fig. \ref{attack}, the proposed scheme achieves the highest ASR and the lowest ASR Std among all schemes for each adversary. 
Thus, the results demonstrate that the proposed scheme outperforms the baselines in terms of both the attack performance and adaptability under varying adversarial scenarios.

\subsection{Ablation Experiment}
Ablation experiments are conducted and presented at both the component and module levels. Specifically, the component-level ablation focuses on the two core components of the proposed scheme, including

$ \bullet \, P1$: DNN based transmit power control,

$\bullet \, P2$:  cGAN based adversarial attack design.\\
The module-level ablation further focuses on the first module of the generator, denoted as $M$, as shown in Fig. \ref{generator}B.

For clarity, the variants of the proposed scheme that retain only $P1$ and only $P2$ are denoted as \textbf{P1} and \textbf{P2}, respectively. The variant obtained by removing $M$ from the proposed scheme is denoted as \textbf{w/o M}. Fig~\ref{ed_} shows the Euclidean distance between the clean and corresponding attacked received signals of \textbf{P1}, \textbf{P2}, \textbf{w/o M} and the proposed scheme under different adversaries, while Fig. \ref{asr-part} presents the ASR and ASR Std of \textbf{P1}, \textbf{P2}, \textbf{w/o M} and the proposed scheme.

\subsubsection{Component-Level Ablation}
For the component-level ablation, we investigate the individual contributions of $P1$ and $P2$ in the proposed scheme. 

As shown in Fig. \ref{asr-part}, \textbf{P1} achieves higher ASR than \textbf{P2}, indicating stronger aggressivity. However, Fig.~\ref{ed_} and Fig.~\ref{asr-part} show that \textbf{P1} also leads to larger Euclidean distance and higher ASR Std than \textbf{P2}, which weakens the undetectability and adaptability of attacks. In contrast, \textbf{P2} produces smaller Euclidean distance compared to that of the proposed scheme in Fig.~\ref{ed_}, as well as lower ASR Std in Fig.~\ref{asr-part}, but it suffers from the reduced ASR.

Thus, the results reveal the complementary effects of $P1$ and $P2$. By jointly incorporating them, the proposed scheme can achieve better performance among aggressivity, undetectability and adaptability, yielding higher ASR, smaller Euclidean distance and lower ASR Std. This validates the necessity of both $P1$ and $P2$ in the proposed scheme.

\subsubsection{Module-Level Ablation}
For the module-level ablation, we further investigate the contribution of $M$ to design the perturbating signals. 

As shown in Fig.~\ref{ed_} and Fig.~\ref{asr-part}, compared with the proposed scheme, \textbf{w/o M} consistently produces lower ASR, larger Euclidean distance and higher ASR Std under different adversaries. This indicates that removing $M$ from the proposed scheme not only weakens the aggressivity, but also makes the generated perturbating signals less undetectable and less adaptive to the dynamic attack channels.

Therefore, $M$ plays an important role in enabling the proposed scheme to achieve better performance in terms of aggressivity, undetectability and adaptability, demonstrating its necessity and effectiveness.

\subsection{Time Consumption Comparison}
\begin{table}[t]
	\centering
	\caption{Comparison of average running time.}
	\label{tab:running_time}
	\begin{tabular}{lc}
		\toprule
		Method & Average Running Time (s) \\
		\midrule
		GP       & $9.6950\times10^{-1}$ \\
		DNN      & $4.4100\times10^{-1}$ \\
		\midrule
		UIAA     & $3.5883\times10^{3}$   \\   
		DLIA     & $2.3630\times10^{1}$   \\   
		FGSA     & $1.5549\times10^{3}$   \\   
		GA       & $5.5100\times10^{-1}$ \\
		CCFA     & $5.8100\times10^{-1}$ \\
		\textbf{Proposed} & $\mathbf{5.7100\times10^{-1}}$ \\
		\bottomrule
	\end{tabular}
\end{table}

Table~\ref{tab:running_time} compares the computational time required by GP solving and the DNN-based scheme to obtain the transmit power solution, as well as the time required by different attack methods to generate the perturbating signals, including CCFA, UIAA, DLIA, FGSA, GA and the proposed scheme. As shown in Table~\ref{tab:running_time}, solving the GP problem once requires more time than performing a single DNN inference, indicating that the DNN-based power control scheme is better suited to the real-time transmit power optimization. Moreover, as shown in Table~\ref{tab:running_time}, UIAA, DLIA and FGSA incur considerably higher computational time than CCFA, GA and the proposed scheme due to the gradient computation, repeated backpropagation and iterative procedure. Although CCFA, GA and the proposed scheme have comparable running time, the proposed scheme achieves superior performance in terms of undetectability, aggressivity and adaptability, thereby demonstrating its overall effectiveness.


\section{Conclusion}
In this paper, we have investigated the adversarial attacks against the DL-based MISO autoencoder communications with multiple receivers and adversaries. To improve the aggressivity, undetectability and adaptability of the adversarial attacks, we propose a DL-based intelligent attack framework consisting of two main components. Specifically, the DNN-based transmit power control is established to mitigate the CLI at each TR by regulating the transmit power of adversaries. This can suppress the CLI caused by the multiple parallel attack links, thereby improving the undetectability. Furthermore, the cGAN-based adversarial attack is developed to generate the perturbating signals considering both the aggressivity and adaptability under the dynamic attack channels. Numerical results demonstrate that the proposed scheme achieves superior performance in terms of aggressivity, undetectability and adaptability.

\setlength{\baselineskip}{12pt}
\bibliographystyle{ieeetr}
\bibliography{references.bib}

\end{document}